\documentstyle[aps,prb,epsfig,preprint,floats]{revtex}

\begin{document}
\draft
\title{Spin-density-wave transition
of (TMTSF)$_2$PF$_6$ at high magnetic fields}
\author{N. Matsunaga, K. Yamashita, H. Kotani and K. Nomura}
\address{Division of Physics, Hokkaido University, Sapporo 060-0810, Japan}
\author{T. Sasaki}
\address{IMR, Tohoku University, Sendai 980-8577, Japan}
\author{T. Hanajiri, J. Yamada, S. Nakatsuji and H. Anzai}
\address{Department of Material Science, Himeji Institute of Technology, Kamigohri 678-1297, Japan}
\date{Submitted Nov. 28, 2000}
\maketitle
\begin{abstract}
The transverse magnetoresistance of the Bechgaard salt (TMTSF)$_2$PF$_6$ has been measured for various pressures, 
with the field up to 24 T parallel to the lowest conductivity direction c$^{\ast}$.
A quadratic behavior is observed in the magnetic field dependence of the spin-density-wave (SDW) transition 
temperature $T_{\rm {SDW}}$.
With increasing pressure, $T_{\rm {SDW}}$ decreases and the coefficient of the quadratic term increases.
These results are consistent with the prediction of the mean-field theory 
based on the nesting of the quasi one-dimensional Fermi surface.
Using a mean field theory, $T_{\rm {SDW}}$ for the perfect nesting case is estimated as about 16 K. 
This means that even at ambient pressure where $T_{\rm {SDW}}$ is 12 K, 
the SDW phase of (TMTSF)$_2$PF$_6$ is substantially suppressed by the two-dimensionality of the system.

\end{abstract}
\narrowtext 

\section{INTRODUCTION}

The Bechgaard salts (TMTSF)$_2X$, where TMTSF denotes tetramethyltetraselenafulvane
and $X$=PF$_6$, AsF$_6$, ClO$_4$ etc., have a rich phase diagram exhibiting spin-density-wave (SDW), metallic, 
superconducting and field-induced SDW (FISDW) phases depending on pressure or anion kind.~\cite{Ishiguro}
The stacking of the TMTSF molecules provides a highly anisotropic Fermi surface (FS) only consisting of open sheets.
For (TMTSF)$_2$PF$_6$, the SDW transition occurs at 12K at ambient pressure. 
With increasing pressure, the SDW phase is suppressed and the superconducting phase is induced at 1.1 K
above the critical pressure of 0.8 GPa.

The phase diagram of the metallic-SDW and FISDW transition for the PF$_6$ salt is well explained 
by the mean field theory based on the nesting of the slightly warped, quasi one-dimensional Fermi surface.
When the magnetic field is applied to the SDW state suppressed by the imperfect nesting of the Fermi surface,
the total energy of the SDW is lowered by the quantization of the closed orbits near the Fermi level.
This corresponds to the fact that the one-dimensional character of electron band is recovered in the magnetic field.
It has been predicted that $T_{\rm {SDW}}$
increases nearly quadratically with the field in low magnetic fields
and shows a saturation behavior to the transition temperature for the perfect nesting case
in high magnetic field.~\cite{Montambaux}
The magnetic field dependence of $T_{\rm {SDW}}$ is quite sensitive to the degree of the one-dimensional character 
of the electron band, i.e. the imperfect nesting of the Fermi surface. 
Accordingly, the behavior of SDW phase in the magnetic field gives important information
about the stabilization mechanism and the characteristic energy of the SDW state. 

In this paper, we present the results of resistivity measurements for the SDW phase
of (TMTSF)$_2$PF$_6$ under pressure and strong magnetic field.
We discuss an applicability of the mean-field predictions and estimate
the characteristic energy of (TMTSF)$_2$PF$_6$ from the data using the mean field theory. 
 
\section{Experiments}

Single crystals of (TMTSF)$_2$PF$_6$ were synthesized by the standard electro-chemical method.
Resistance measurements along the conducting a-axis
 were carried out using a standard four probe dc method. 
 The typical size of the sample was 1$\times$0.1$\times$0.1 mm$^3$.
Electric leads of 10 $\mu$m gold wire were attached with silver paint onto 
gold evaporated contacts.
The current contacts covered the whole areas of both ends of the sample for an uniform current.
The temperature was measured using a Cernox CX-1050-SD
resistance thermometer calibrated by a capacitance sensor in magnetic fields.
The sample was mounted inside a beryllium-copper clamp cell with Daphne 7373 oil as a pressure medium.
It is known that the pressure decreases with decreasing temperature and the pressure reduction from 300 K
to 4.2 K is approximately 0.15 GPa, irrespective of the initial pressure at 300 K
in the present experimental configuration.~\cite{Murata}
The measurements reported here were carried out on two samples in fields to 24 T
(sample $\#1$) in a hybrid magnet at the High Magnetic Field Laboratory, IMR, Tohoku University
and in fields to 16 T (sample $\#2$) in a superconducting magnet at Hokkaido University.

\section{RESULTS AND Discussion}

\begin{figure}
\epsfxsize=3.0in \center \epsfbox{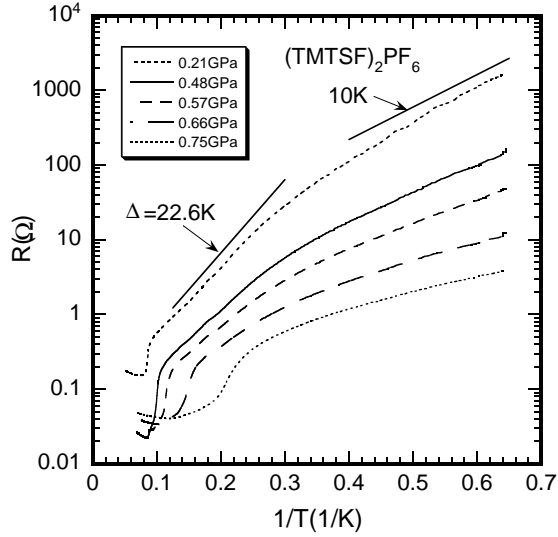} \vspace{0.1in}
\caption{Temperature dependence of the transverse resistance in (TMTSF)$_2$PF$_6$ for various pressures. 
Data are for sample \#2. The straight lines show the apparent activation behavior.
} \label{fig:5}
\end{figure}

The temperature dependence of the transverse resistance in (TMTSF)$_2$PF$_6$ for various pressures is shown 
in Fig. 1.
For every pressure, the resistance shows a steep increase associated with the metal-SDW transition.
The SDW transition temperature $T_{\rm SDW}$ is determined from the peak position
in the derivative of the logarithm of these data with 1/$T$. 
The peak height at $T_{\rm SDW}$ decreases with increasing pressure.
Although the transition width, which is estimated from the peak width in the derivative curve,
is broadened with decreasing $T_{\rm SDW}$, the peak is fairly sharp at every pressure.
The temperature and pressure dependence of the resistance are consistent with previous results.~\cite{Biskup}
More detailed behaviors of the resistance will be discussed later.

\begin{figure}
\epsfxsize=3.0 in \center \epsfbox{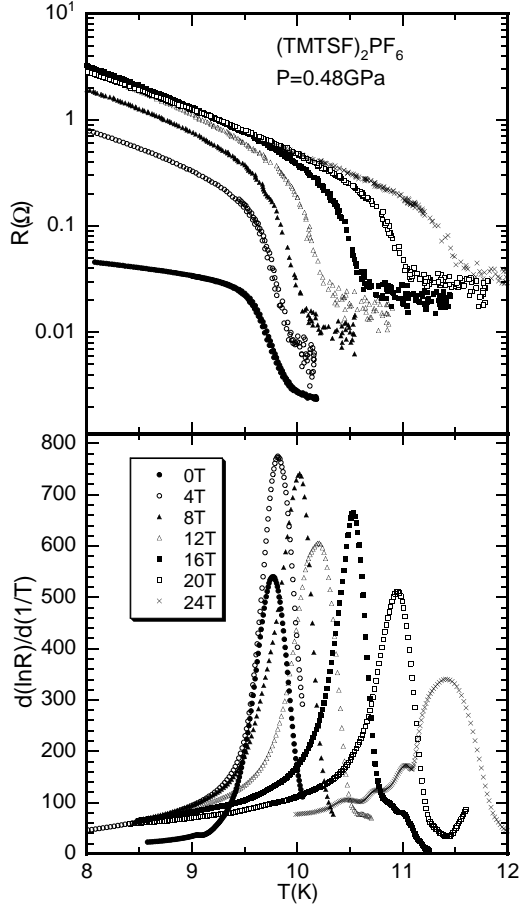} \vspace{0.1in}
\caption{(Top) Temperature dependence of the transverse magnetoresistance in
(TMTSF)$_2$PF$_6$ at 0.48 GPa in a constant field along the $c^{\ast}$ axis as indicated.
(Bottom) Logarithmic derivatives of the data.
Data are for sample \#1.} \label{fig:1}
\end{figure}

Figure 2 shows the temperature dependence of the transverse magnetoresistance with the field $B$ parallel 
to the lowest conductivity direction $c^{\ast}$ at 0.48 GPa.
In each field, the resistance shows a steep increase, associated with the SDW transition,
with decreasing temperature. 
From this figure, it is clear that the transition is shifted 
towards higher temperatures as the field is increased.
We determine $T_{\rm SDW}$ from the peak in the derivative of 
the logarithm of the resistance with 1/$T$, which is also shown in Fig. 2.
The field dependence of $T_{\rm {SDW}}$ is shown in Fig. 3.
At 0.48 GPa, $T_{\rm {SDW}}$ increases from 9.8 K in zero field up to 11.4 K at 24 T.
This field dependence of $T_{\rm {SDW}}$ is well described by the quadratic behavior as 
\begin{equation}
T_{\rm {SDW}}(B)=T_{\rm {SDW}}(0)+CB^2,
\end{equation}
with a constan $C$.
The solid line in Fig.3 is the best fit of the quadratic function to the experimental data.
The observed result for $T_{\rm {SDW}}$ is well described by a quadratic behavior 
and there is no saturation even at 24 T.

Figure 4 shows the magnetic field dependence of $T_{\rm {SDW}}$ for various pressures.
With increasing pressure, $T_{\rm {SDW}}$ decreases and the coefficient $C$ of the quadratic term increases.
These results are qualitatively consistent with the prediction of the mean field theory.~\cite{Montambaux}
The solid line in Fig. 4 is also the best fit of the quadratic function to experimental data.
We obtain the coefficient $C$ from the fitting 
and summarize the relation between $T_{\rm SDW}$ and the coefficient $C$ in log-log scale in Fig.5. 
We also plot the earlier data for (TMTSF)$_2$PF$_6$ for comparison\cite{Danner,Kwaket,Uji}.
There is good agreement between the present result and previous ones, 
although a slight scatter is seen in the high pressure.

\begin{figure}
\epsfxsize=3.0in \center \epsfbox{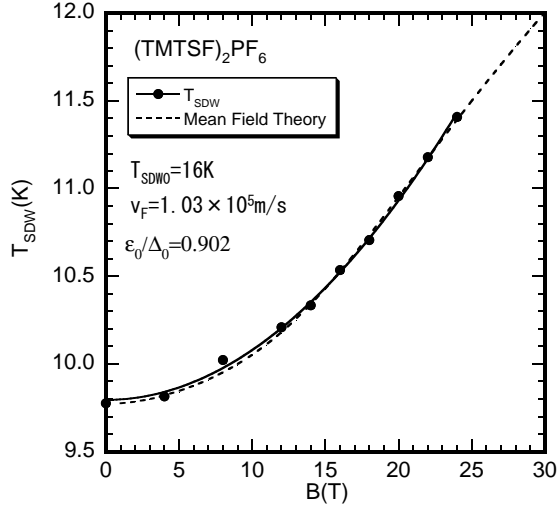} \vspace{0.1in}
\caption{Magnetic field dependence of the SDW transition temperature $T_{\rm {SDW}}$ at 0.48 GPa.
The full line is the best fit of the quadratic function.
The dashed line is the fit using mean field theory.} \label{fig:2}
\end{figure}

From the observed curvature of the coefficient $C$ of the quadratic term as a function of $T_{\rm {SDW}}$ 
for various pressures, we can determine several important parameters 
for the SDW transition in a mean-field theory.
The dashed line in Fig. 5 is the fit of this theory to the experimental data under low pressure and
the agreement is good.
The fit gives $T_{\rm SDW}$ corresponding 
to the perfect nesting case as $T_{\rm {SDW_0}}$=16 K and Fermi velocity $v_{\rm F}$=1.03$\times$10$^5$ m/s, 
with a lattice parameter along the $b^{\prime}$ axis of 7.7$\times$10$^{-10}$ m.
From these parameters, we can determine the imperfect nesting parameter $\epsilon_0$/$\Delta_0$ 
where $\epsilon_0$ is the energy parameter characterized the deviation from perfect nesting and
$\Delta_0$ is the SDW order parameter at T=0K for $\epsilon_0$=0.
At ambient pressure where $T_{\rm {SDW}}$ is 12 K, we obtain $\epsilon_0$/$\Delta_0$=0.8 for (TMTSF)$_2$PF$_6$.
Although the value of $\epsilon_0$/$\Delta_0$ is somewhat large in comparison 
with previous estimates \cite{Biskup},
this parameter is consistent with the value estimated from the analysis of the STM spectroscopy.~\cite{Ichimura}
We summarize the value of $T_{\rm SDW}$ and $\epsilon_0$/$\Delta_0$ obtained from this fit for various pressures
in Table 1.
It should be noted that even at ambient pressure where $T_{\rm {SDW}}$ is 12 K, 
the SDW phase of (TMTSF)$_2$PF$_6$ is substantially suppressed by the two-dimensionality of the system.

According to a more general description for the SDW 
in the magnetic field based on the mean-field theory~\cite{Bje,Maki}, 
the magnetic field dependence of $T_{\rm SDW}$ is described by
\begin{eqnarray}
\lefteqn{\ln{\left[ {T_{\rm {SDW}}(B) \over T_{\rm {SDW_0}}} \right]} ~\simeq~}\hspace{7cm} \nonumber \\
\sum_{l=-\infty}^{\infty} J^2_l \left( {\varepsilon_0 \over \omega_b} \right)
\left\{ Re\Psi({1 \over 2}+{il\omega_{\rm b} \over 2\pi T_{\rm {SDW}}})-\Psi({1 \over 2}) \right\},
\end{eqnarray}
where $J_l$ and $\Psi$ are Bessel and digamma functions respectively, 
$\omega_{\rm b}=v_{\rm F}eb^{\prime}B$ is the cyclotron frequency along the b-axis and e is the electron charge.
The dashed line in Fig. 3 is the curve calculated from Eq. (2) using the above parameters.
The experimental results are well reproduced by the mean field calculation,
indicating that the values obtained for these parameters are reasonable.
The mean field calculations nicely account for the field induced increase of 
$T_{\rm {SDW}}$ and explain the absence of saturation behavior in the present experimental range of 
magnetic field as shown in Fig. 3. 
Although the value of Fermi velocity is somewhat small in comparison 
with a previous estimate $v_{\rm F}$=1.5$\times$10$^5$ m/s from the resistance measurements assumed 
that (TMTSF)$_2$PF$_6$ at ambient pressure was the perfect nesting case \cite{Biskup},
these parameters are consistent with the fact that there is no deviation
from the quadratic behavior of $T_{\rm {SDW}}$ up to 28 T at ambient pressure \cite{Danner,Uji}.

\begin{figure}
\epsfxsize=3.0in \center \epsfbox{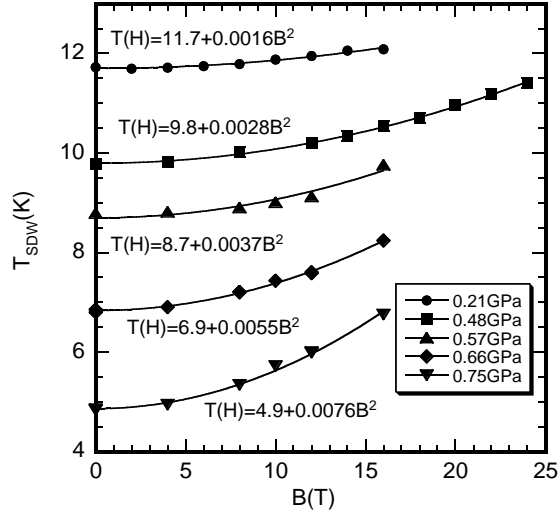} \vspace{0.1in}
\caption{Magnetic field dependence of the SDW transition temperature $T_{\rm {SDW}}$ for various pressures.
The full line is the best fit of the quadratic function. } \label{fig:3}
\end{figure}

\begin{figure}
\epsfxsize=3.0in \center \epsfbox{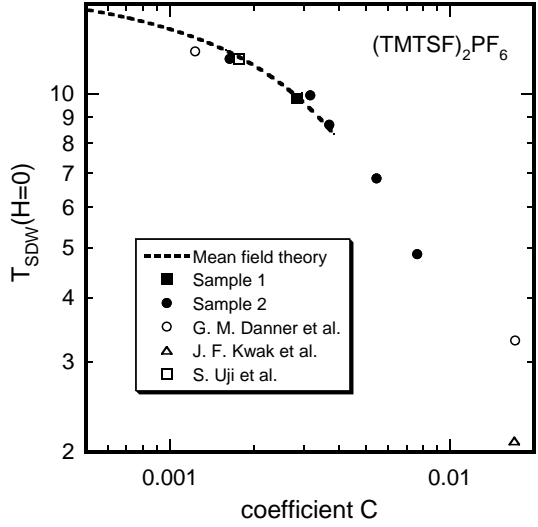} \vspace{0.1in}
\caption{$T_{\rm {SDW}}$ vs coefficient $C$ of a quadratic term for (TMTSF)$_2$PF$_6$, 
together with $C$ for previous results \cite{Danner,Kwaket}.
The dashed line is the fit of mean field theory.
} \label{fig:4}
\end{figure}

\begin{figure}
\epsfxsize=3.0in \center \epsfbox{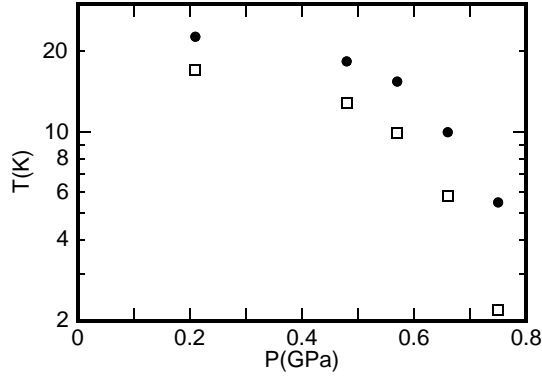} \vspace{0.1in}
\caption{Pressure dependence of the apparent activation energy at 0.5$T_{\rm SDW}$ ($\bullet$) and 
the geometrically averaged activation energy $\Delta_G$ ($\Box$) (see text).
} \label{fig:5}
\end{figure}

\begin{table}
\caption{SDW transition temperature $T_{\rm {SDW}}$, imperfect nesting parameter $\epsilon_0$/$\Delta_0$ 
at five different pressures.}
\begin{tabular}{ccc}

P(GPa)&$T_{\rm {SDW}}$(K)&$\epsilon_0$/$\Delta_0$\\ \tableline
0.21&11.7&0.794 \\
0.48&10.0&0.889 \\
0.57& 8.8&0.935 \\
0.66& 6.9&0.979 \\
0.75& 4.9&0.997 \\ 
\end{tabular}
\end{table}

Finally, we discuss the temperature and pressure dependence of the resistance.
As shown in Fig.1, the transverse resistance roughly shows an activation-type temperature dependence
below $T_{\rm SDW}$, however, the situation is much more complex than in the usual activated insulator.
The slope of the resistance against 1/$T$, that corresponds to the apparent activation energy, 
slowly decreases from $\Delta$=22.6 K at 0.5$T_{\rm SDW}$ to $\Delta$=10.0 K below 2 K under 0.21 GPa.
These results are consistent with previous results.~\cite{Uji}
Usually, one would expect that the activation energy is simply given by the minimum value of the gap,
as $\Delta_M= \Delta_0-\varepsilon_0$.
With $T_{\rm {SDW_0}}$ =16K, $\epsilon_0$/$\Delta_0$=0.794 
and the mean-field BCS relation for the weak-coupling limit as 2$\Delta_0$ = 3.5$k_{\rm B}T_{\rm SDW_0}$,
we obtain the direct gap 2$\Delta_0$ = 56 K and $\Delta_M$ = 5.6 K for 0.21 GPa.
Although the value of 2$\Delta_0$ is somewhat small in comparison 
with a previous estimate from an optical measurement, where 2$\Delta_0$ = 101 K \cite{Mihaly,Vescoli},
this value is consistent with the value estimated from the analysis of the STM spectroscopy,
where 2$\Delta_0$ = 58$\sim$67 K.~\cite{Ichimura}
The value of $\Delta_M$ is much smaller than the obtained value from the resistance at 0.5$T_{\rm SDW}$.
The activation-type temperature dependence with the minimum value of the gap is correct only well below 
the temperature corresponding to the activation energy $\Delta_M$.
Above and around this temperature,
the temperature dependence of the resistance is better defined by considering the geometrically averaged 
activation energy of the state density\cite{Maki} as
\begin{equation}
\Delta_G=\sqrt{\Delta_0^2-\varepsilon_0^2}.
\end{equation}
From this definition, we obtain the geometrically averaged activation energy $\Delta_G$=17 K for 0.21 GPa.
This value is close to the value of the apparent activation energy
around 0.5$T_{\rm SDW}$.
The slow decrease of the apparent activation energy with decreasing temperature can be explained by 
the crossover from the geometrically averaged activation energy $\Delta_G$ to the activation energy $\Delta_M$.
With increasing pressure, the apparent activation energy $\Delta$ at 0.5$T_{\rm SDW}$ 
decreases from 22.6 K for 0.21 GPa to 5.5 K for 0.75 GPa.
These results are also consistent with previous results.~\cite{Biskup}
With increasing pressure, the ratio of the apparent activation energy at 0.5$T_{\rm SDW}$ to 
$\Delta_G$ becomes large as shown in Fig.6.
According to the mean field theory, the temperature dependence of order parameter becomes 
stronger with increasing two-dimensionality of the electron band and an increase of order parameter
with decreasing temperature gives a steep decrease of the number of electrons activated 
across the gap.~\cite{Yamaji}
As a result, the apparent activation energy modified by the temperature dependence of the order parameter increases 
with increasing pressure and the pressure dependence of the ratio of the apparent activation energy 
at 0.5$T_{\rm SDW}$ to $\Delta_G$ can be explained by this temperature dependence of the order parameter.
Finally, the temperature dependence of the resistance below $T_{\rm SDW}$ 
is successfully explained by the mean-field theory with the same parameters as the discussion for $T_{\rm SDW}$.

\section{Conclutsion}

We have measured the resistivity in the SDW phase
of (TMTSF)$_2$PF$_6$ under various pressures and magnetic fields parallel to the c*-axis. 
The SDW transition temperature in zero field decreases with increasing pressure. 
When the magnetic field is applied to this SDW phase suppressed by the pressure,
$T_{\rm {SDW}}$ recovers its value.
The field dependence of $T_{\rm {SDW}}$ is described by a quadratic behavior 
and the coefficient of the quadratic term increases with increasing pressure. 
These results are consistent with the prediction of the mean field theory based on nesting of the Fermi surface.
We have determined the SDW transition temperature for the perfect nesting case as $T_{\rm {SDW_0}}$ =16 K 
and the imperfect nesting parameter $\epsilon_0$/$\Delta_0$=0.8 for (TMTSF)$_2$PF$_6$ at ambient pressure.
This means that the SDW phase of (TMTSF)$_2$PF$_6$ at ambient pressure, whose $T_{\rm {SDW}}$ is 12 K,
is suppressed by the two-dimensionality of the system.
The temperature and pressure dependence of the activation energy in the SDW phase of (TMTSF)$_2$PF$_6$
is also well explained by the mean-field theory using the same characteristic energy in the SDW state
by taking into account the geometric average of the state density.

\acknowledgments
We would like to thank Professor K. Murata for technical advice on the high pressure measurement.
Some of this work was carried out as part of the
"Research for the Future" project, JSPS-RFTF97P00105,
supported by the Japan Society for the Promotion 
of Science.

\end{document}